\documentclass[12pt]{article}
\input{epsf.tex}
\begin{document}

\begin{center}
{\bf PARAMETERS OF THE BEST APPROXIMATION FOR DISTRIBUTION OF THE REDUCED
NEUTRON WIDTHS. SPECIFICITY OF FULL-SCALE METHOD OF ANALYSIS}\\
\end{center}
\begin{center}
{\bf  A.M. Sukhovoj, V.A. Khitrov}\\
\end{center}
\begin{center}
{\it Joint Institute for Nuclear Research, Dubna, Russia}
\end{center}
\begin{abstract}

The method is described and tested for analysis of statistical parameters
of reduced neutron widths distributions accounting for possibility
of coexistence of superposition of some functions with non-zero mean
values of neutron amplitude and its arbitrary dispersion.
The possibility to obtain reliable values of distribution parameters at
variation of number of resonances involved in analysis and change of
registration threshold of resonances with the lowest widths is studied.
\end{abstract}

\section{Introduction}\hspace*{16pt}

Experimentally measured reduced widths $\Gamma_n^0$ ($\Gamma_n^1$) of neutron
resonances  -- strongly fluctuating values.
This circumstance very much complicates determination of their mean values
(the averaged spacing $D$ and strength function  $S_0=<\Gamma_n^0>/D_0$)
from real experimental data distorted by different systematical uncertainties.
The generally accepted notion of shape of their distribution was suggested
in 1956 \cite{PT} and was not up to now tested in full scale.  

This test is non-trivial procedure because only the part of the measured
distribution is observed in experiment but it's independent parameter $X$
can be determined only for the total spectrum of possible values
of widths. Id est, approximation of experimental data is performed at presence
of unknown error parameter $X = \Gamma_n^0 /<\Gamma_n^0>$.
Real value and error $<\Gamma_n^0>$, of cause, cannot be determined
experimentally. And value of the  $\delta X$ depends on accepted model notions.

The ordinary test of distribution $\Gamma_n^0$ consists in determination
of effective value of number of degrees of freedom $\nu$ of
$\chi^2$-distribution for given set of widths (with fixed orbital momentum  $l$)
and rather subjective choice of neutron energy interval where distortions of
$D$ and $S$ are minimal.
The deeper test must answer the questions, in what degree are realized the
conditions of applicability of $\chi^2$-distribution to real data. Id est:

(a) whether mathematical expectation of amplitude $A=\sqrt{\Gamma_n^0}$ is
equal to zero,

(b) its dispersion -- to mean value  $<\Gamma_n^0>$ and

(c) whether the function providing description of experimental data with
maximum possible precision is the unique?

It should be also taken into account that practical investigation of nucleus
properties includes obligatory stage -- creation of mathematical model of
process under study. By this, any model is created on limited basis of data
having non-estimated systematical errors which are inevitably projected on
the following investigation of nucleus properties.

Therefore, predictive ability of a model and its quality are strongly
correlated values. It follows from this the necessity to test hypothesis
\cite{PT} Ц whether real distribution $A$ can be composition of several Gauss
distributions with different mean values and dispersions.

\section{Choice of experimental data presentation form}\hspace*{16pt}

Analysis of status of the problem from the point of view of both theoretical
ideas and totality of experimental data allows one to expect for maximal
discrepancy between experimental data and hypothesis \cite{PT} in region
of maximal widths. In practice, it can be caused by influence
\cite{PEPAN-1972} of large components of wave functions of nuclear
states with maximal number of quasi-particles and phonons owing to
their weak fragmentation \cite{MalSol} in the excitation energy region
$E_{\rm ex} \approx B_n$.

Distribution $\Gamma^0_n$ can be approximated in both its ``differential"
and ``integral" forms in function of width or square root from this resonance
parameter.  Experimental data contain fixed quantity of information.
Therefore, the volume of available information does not depend on form of
data presentation and its choice is determined only by mathematical problems
of obtaining of the sought values and visualization of results.
In principle, it is possible to ahalyse both distribution itself or large enough set
of its momentums.

Specific problem of distribution analysis of changeable values at presence
of threshold of their registration  - the lack of information on portion
of distribution of neutron width is really observed in experiment.
As a consequence, there appears the problem of unit of measurement for
random value Ц it does not depend on form of distribution presentation.
The most suitable form for presentation of the data for the  problem under
solution is cumulative sum of experimental values of $X=\Gamma^0_n/<\Gamma^0_n>$,
increasing when increases $X$. This sum includes all the observed
experimentally and included in the used compilation (for example, in \cite{BNL}
or library ENDF/B-VII \cite{IAEA}) values of widths. 

The selective average $<\Gamma^0_n>$ for experimental cumulative sum was determined
from this set without accounting for missed resonances and their unresolved
multiplets. Its inevitable displacement with respect to unknown value is
compensated at approximation by deflection of approximated $\sigma$ value from
the most probable value (in particularly, from $\sigma=1$). 
This uncertainty does not influence $\chi^2$ - the shape of relative difference
between experimental and approximated distributions does not depend on units
determining the width $\Gamma^0_n$.

Approximation region in all calculations was limited by the interval from
zero to twice maximal experimental values $X_{\rm max}$. Cumulative sum was
normalized in point $X_{\rm max}$ to number of experimentally determined widths.
The region $(0-2X_{\rm max})$ included in all cases not less
than 1000 points, in which was minimized the difference of experimental
cumulative sum and its approximating function. Dispersion of cumulative
sum at this normalization changes from zero in extreme points to maximal value
in region $X \sim 2-10$ (see Fig. 1). In given variant of analysis this
change was ignored, and $\chi^2$ was calculated as a sum of squares of
difference of experimental and approximated values of cumulative sums.
Naturally, all statistical errors in region of the lowest widths in this
case exert the lowest influence on determined parameters of distributions.
For convenience of comparison of different data the value $\chi^2$ was
divided by number of freedom degrees of approximation.

All the obtained experimentally values of widths were included in
practical analysis except obvious errors of experiment
(misprints in compilation, strong discrepancy in different data sets).
Practically, the latter can be revealed only in region of maximal
values of $\Gamma^0_n$, therefore corresponding  correction decreases
degree of discrepancy between experiment and \cite{PT}.

This form of presentation of experimental data permits one to involve
simply enough in approximation, in principle, any factor distorting
width distributions. Besides, this allows determination of probable resonance
parameters for any nucleus at presence of systematical errors of $\Gamma^0_n$,
if only influence of such systematical error can be take into account in any
(numerical or analytical) form of functional dependence with free parameters.

\section{Model and method of suggested analysis}\hspace*{16pt}

Experimental level density in region of neutron resonances of nuclei from
mass region $40 \leq A \leq 200$ obtained in Dubna (within the model-free
method for analysis of the two-step cascade intensities) is described
\cite{Prep196, PEPAN-2006, PHAN7210} by sum of three (or more) partial
level densities with different number of quasi-particles and phonons.
Practically, it was accepted on calculation problems that in limit case the
experimentally observed resonances can belong (as a maximum) to four
different distributions of $\Gamma^0_n$ for even-even target-nuclei.
This is true and for $A$ odd nuclei at equality of $<2g\Gamma^0_n>$ for
resonances with different spins $J$. In the other case the results of
approximation contain and information on spin dependence of neutron
strength functions.

Physically, according to the parameters of different approximation
variants of the total set of level density obtained for $\approx$40 nuclei
in Dubna, it is also worth while to limit maximal  value of $K$ by $K=4$.
In this case, the system of corresponding nonlinear equations will be
most probably always degenerated. Therefore, instead of determination
of the unique value of any parameter, it is necessary and possibly to
determine the width of limited interval of their values corresponding
to $\chi^2$ minimum.

A smallness of set of experimental values of the widths and exponential
functional dependence of probability for their observation at different
$\Gamma^0_n$ very strongly complicate process of determination of the
parameters for approximating function. Therefore, it is worth while to
perform this operation so that the algorithm of search for minimum of
$\chi^2$ would permit stable approximation of the experimental data at
presence of two and more distributions with practically coinciding parameters.
Comparison of the data obtained for $K>1$ with the results of their
approximation by the only distribution can give new information on nuclear
structure in region $B_n$. First of all -- information on possible
existence of neutron resonances with different structures of their wave
functions (as it was suggested in \cite{Appr-k}).

Practical degeneration of the realized process together with exponential
change of the analyzed dependencies complicate (but do not exclude)
the use of the Gauss method for solution of systems of nonlinear equations
in form of existing library programs. The problem of the use of this method
is very complicated by events of appearance (as the most probable value) of
near to zero values $\sigma$ and corresponding to them steps in cumulative sums.
Id est -- to
some sets of non-random  width values.
 There is easier realized the Monte-Carlo method for solution of systems of
degenerated nonlinear equations. Namely -- random set of elements
correction vector of parameters of fitted function with arbitrary variation
of their initial values.

The fitted function is sum $K$ of the distributions $P(X)$ of normally
distributed random values with independent variables  $X_k$ each. 
The required parameters in compared variants are the most probable value
$b_k$ of the amplitude $A=\sqrt{\Gamma_n^0/<\Gamma_n^0>}$,
its dispersion $\sigma_k$  and total contribution  $C_k$ of function number
$k$ for variable 
\begin{equation}
X_k=((A_k-b_k)^2)/\sigma^2_k
\end{equation}
in the total experimental cumulative sum of widths.

The number of distribution and sign of amplitude $A_k$ for given resonance
are unknown. Further was used its positive value because (1) is
invariant with respect to simultaneous change of signs of $A_k$ and $b_k$.
But, it was everywhere supposed that in the considered distribution number
$K$ can exist the only value of $b_k$. I.e., any distribution of widths
$K$ has only one maximally possible value of amplitude. There is the main
(and absolutely necessary) hypothesis of the performed by us analysis of
distributions of the resonance reduced widths. 
Concrete value of function $P(X)$ for  variable (1) in the described analysis
was obtained by compression and shifting of the generally known Euler
gamma-function. The obtained in this way value corresponds to the magnitude
of  this mention function for the variable $X=(A \times \sigma +b)^2$.
At present the basis for this algorithm for setting of parameters of
approximating function is excellent degree of description of all
known experimental distributions of the widths.
In addition it should be
noted that the modules of the $b_k$ and  $\sigma_k$ values are strongly
correlated variables, at least, for large enough  $b_k$ values.


\section{Results of test of analysis method}\hspace*{16pt}

The test of analysis method was performed by approximation of different sets
of random $X$ values. By this, the mean values of normally distributed random
values, their dispersion, number of variables in sets and distortions of different
types appearing in the experiment are easily varied. 

The random value $X=\xi^2$, corresponding to the $\chi^2$- distribution with one
degree of freedom and unit dispersion and corresponding average was generated
from the normally distributed random values $\xi$. The later were set using
Neumann algorithm as a product of two random numbers: $\delta_1=sin(2\pi\gamma)$
and  $\delta_2=-2 ln(\gamma)$, where $\gamma$ -- the uniformly distributed in
interval [0,1] random value. Modeling of experimental distortions of widths
in the case under consideration reduces to corresponding arithmetic
operations with $\xi$ and approximation of cumulative sums of the
distorted values $X_d$ with necessary repetition number of this process.
For example, below modeling of influence of the observation threshold of
resonance was done using linear function of number $X$ with the parameters
providing in sum exclusion from the tested set $L=30\%$ of the lowest
random $X$ values. The spectrum of possible values of cumulative sums
for any practically achievable values of number of observed resonances
can be obtained by interpolation of the data presented in Fig. 1.
 \begin{figure}\begin{center}

\vspace{4cm}
\leavevmode
\epsfxsize=15cm 
\epsfbox{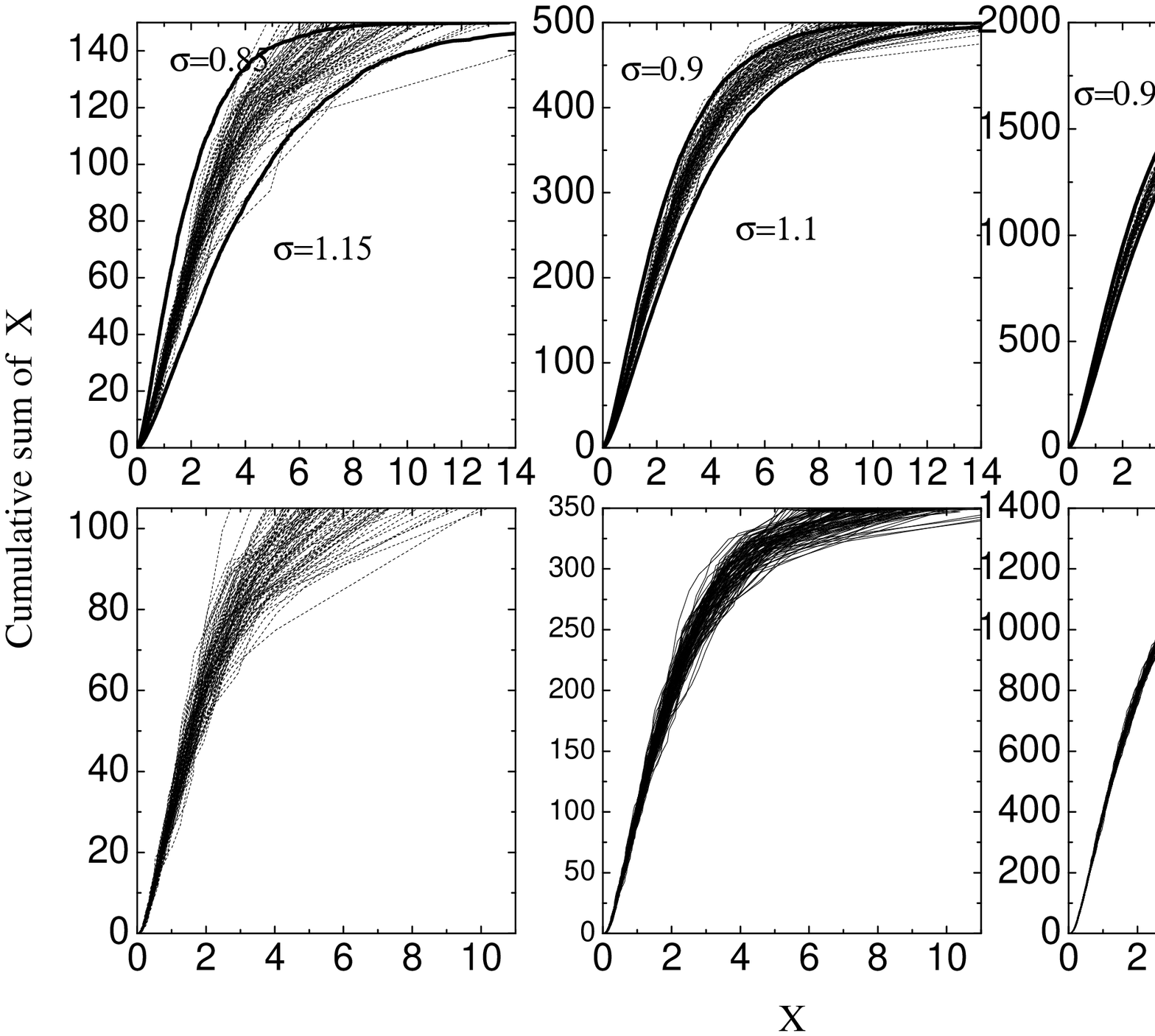} 
\end{center}
\vspace{-5.5cm}
Fig. 1. Thin lines -- the example of cumulative sums for some tens of
sets from 150, 500 and 2000 random $X$ values (upper row)
Thick lines -- the minimal and maximal values with corresponding parameters $\sigma$.
Cumulative sums for the same sets after exclusion of 30\% of the lowest $X$ values
(lower row).

\end{figure}

Large dispersion of random values $X$ brings to large fluctuations of cumulative
sums of both experimental data and model distributions. And, respectively,
to essential variations of the best values of the parameters (1).
Therefore, the conclusion about possible deviations of the parameters $b$ and
$\sigma$ from the  expected values 0 and 1, respectively, can have only
probabilistic character.  

Frequency distributions of these parameters were obtained from modeling sets
for the N=150, 500 and 2000 random values $X$. Modeling was performed for
variant of the non-distorted values $X$ and omission which corresponds to
exclusion of $L=30\%$ of their lowest values (linearly changing with number
of random value). The results of approximation of these distributions
(corresponding to practical maximum of $\chi^2$) are given in Fig. 2.

 \begin{figure}\begin{center}

\vspace{4cm}
\leavevmode
\epsfxsize=15cm

\epsfbox{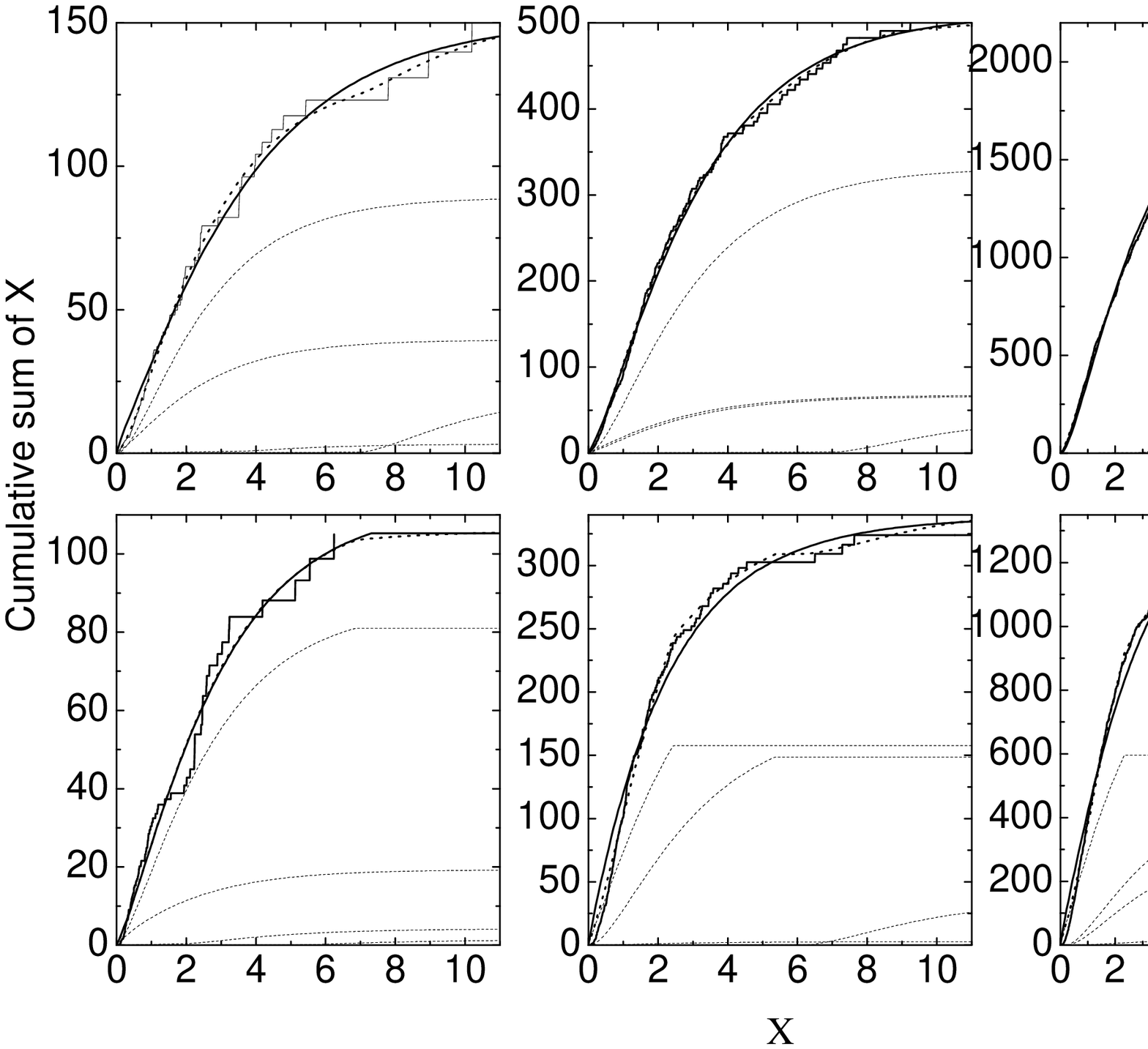} 
\end{center}
\vspace{-5.5 cm}

Fig. 2. The examples of approximation of cumulative sums from the sets presented
in Fig.1 curves for case of maximal $\chi^2$ values. The upper row -- for zero threshold,
the lower row --  with exclusion of 30\% of the least random $X$ values.
Dotted curves -- partial distributions for $K=4$, points -- their sum,
solid curves -- approximation for $K=1$. 

\end{figure}

The widths of corresponding distributions decrease as $N$ increases and at
small $X$ depend on value $L$. One can conclude from the data presented in
figures 1 and 2 that a deviation of the experimental distribution of widths
from the Porter-Thomas distribution appears itself mainly at
$X=(\Gamma^0_n/<\Gamma^0_n>)>2-5$.

Discrepancy between the experimental data and hypothesis at smaller $X$
values can be related, first of all, with omission of weak resonances or
other systematical errors of the experiment. But, it is not excluded and
possibility of real deviation of parameters $b$ and $\sigma$ from values
corresponding to hypothesis \cite{PT}.
 
Probabilistic conclusions on this account can be made only from comparison
between frequency distribution of the parameters (1) for different model
distributions and experimental data. For the case $b=0$ and $\sigma=1$ they
are shown in figures 3 and 4.

 \begin{figure}\begin{center}

\vspace{3cm}
\leavevmode
\epsfxsize=12.5cm

\epsfbox{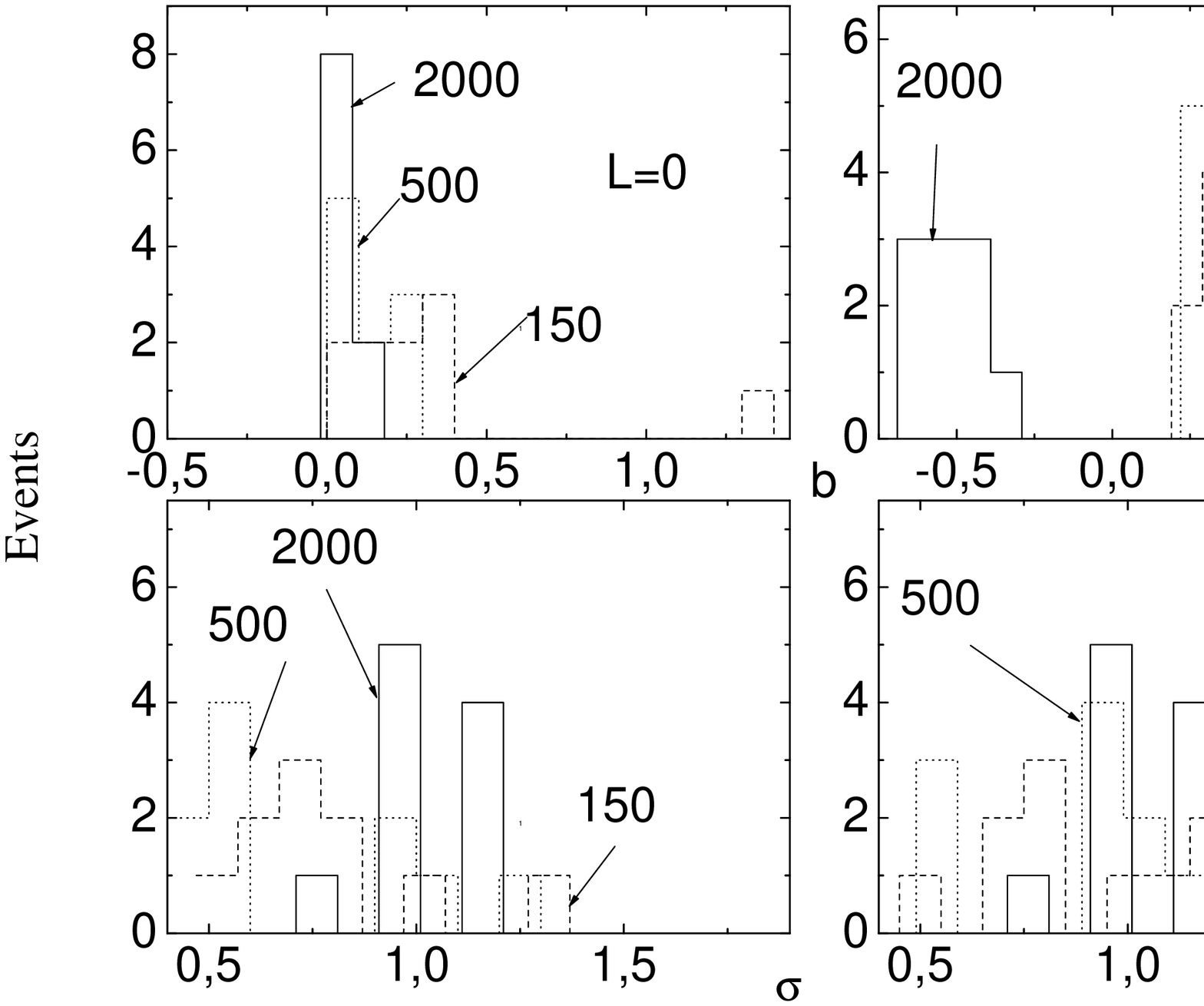} 
\end{center}
\vspace{-4.5cm}

 Fig. 3. Comparison of frequency distributions of given values of $b$
  (upper) and $\sigma$ (lower) rows, respectively for $K=1$.
  Left column -- all possible random values are included in modeling,
  right column -- $L=30\%$ of the lowest random values are excluded
  from each tested set.

\end{figure}

 \begin{figure}[tbp]
\begin{center}

\vspace{3cm}
\leavevmode
\epsfxsize=12.5cm

\epsfbox{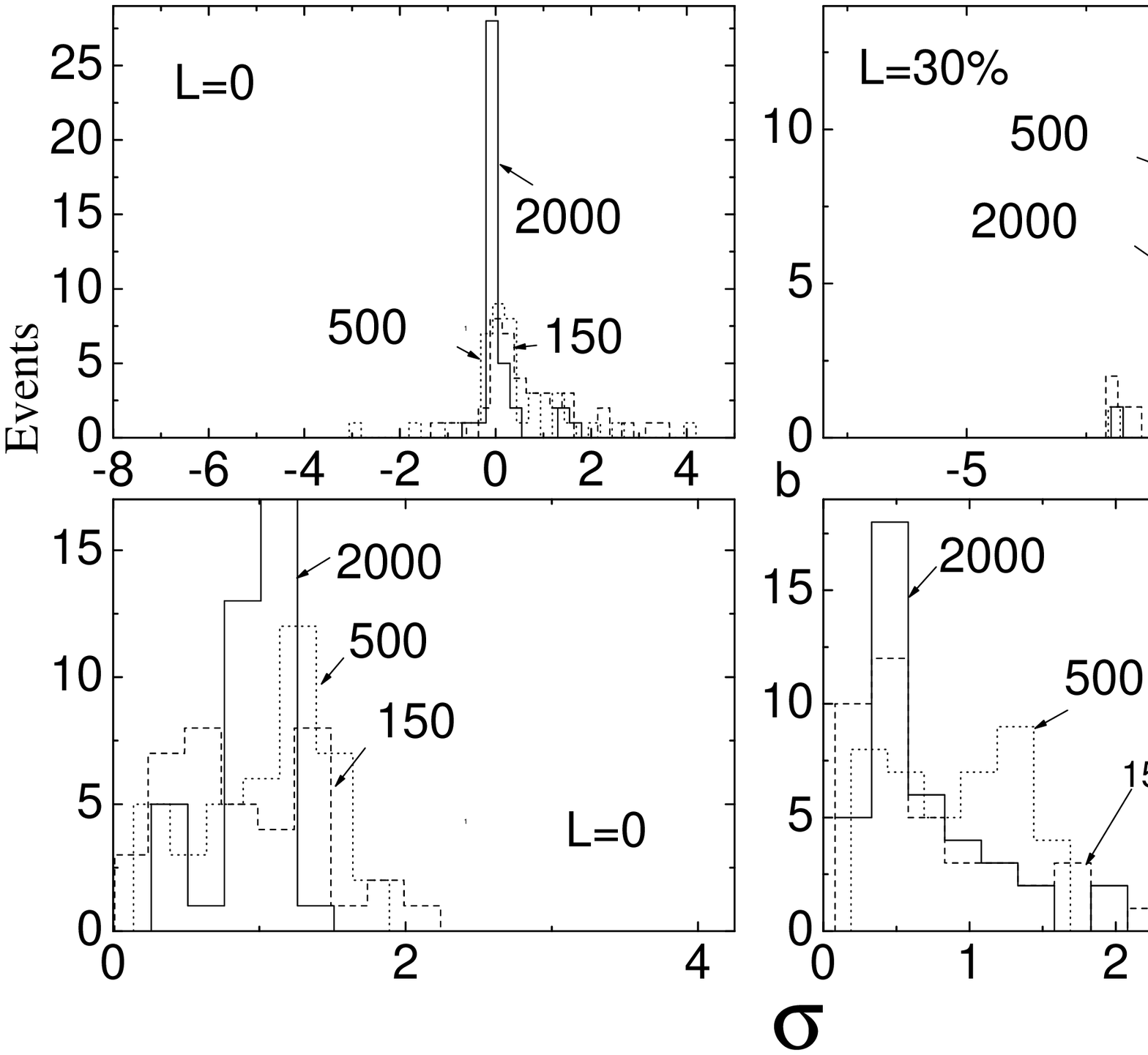} 
\end{center}
\vspace{-4.5cm}
Fig. 4. The same, as in Fig. 3, for $K=4$.   
\end{figure}

\section{Test of method for determination of number of unobserved resonances}\hspace*{16pt}

Any errors of experimental values of the tested set inevitably increase
dispersion of the obtained best values of $b$ and $\sigma$.
But, in principle they can be taken into account by determination of the most
probable parameters even for distorted distribution $\Gamma_n^0$.
For example, the problem of resonance omission can be solved easily enough at
presence of reliably established dependence of threshold of its registration on
neutron energy. A possibility to realize of this computational process follows
from the data presented in Fig. 5. As it is seen from comparison of mean values
of cumulative sums, their form for the same number of resonances
(in given case $N_{exp}$=350) depends on presence/lack of omitted resonances.

Direct estimation of the most probable number of omitted resonances in any
experiment does not call troubles and can be simple if only functional
dependence of their portion $\delta \psi_{th}$ from the total number $S$ is set
on the ground of some data or hypotheses for concrete intervals of
resonance energies. Then
 
\begin{equation}
\chi^2=(S-\psi(A,b,\sigma)-\delta \psi_{th})^2
\end{equation}

Here  $\psi(A,b,\sigma)=\int{X*P(X)dX}$ for any fitted distribution $P$ in
function of ratio $X$. The value $\delta \psi_{th}$ depends only on difference
of $N_{t}-N_{exp}$ for varied from variant to variant expected number of
resonances $N_{t}$ in interval $\delta E$ and determined in the experiment $N_{exp}$.
A number of these intervals was varied in interval 5-20 in dependence on
bulk of the experimental width values.  Moreover, negative values
$N_{t}-N_{exp}$ in any intervals were changed by zero.
The desired value $D=\sum \delta E/\sum N_t$ corresponds to minimum of $\chi^2$.
  
 \begin{figure}
\begin{center}
\vspace{3cm}
\leavevmode
\epsfxsize=12cm

\epsfbox{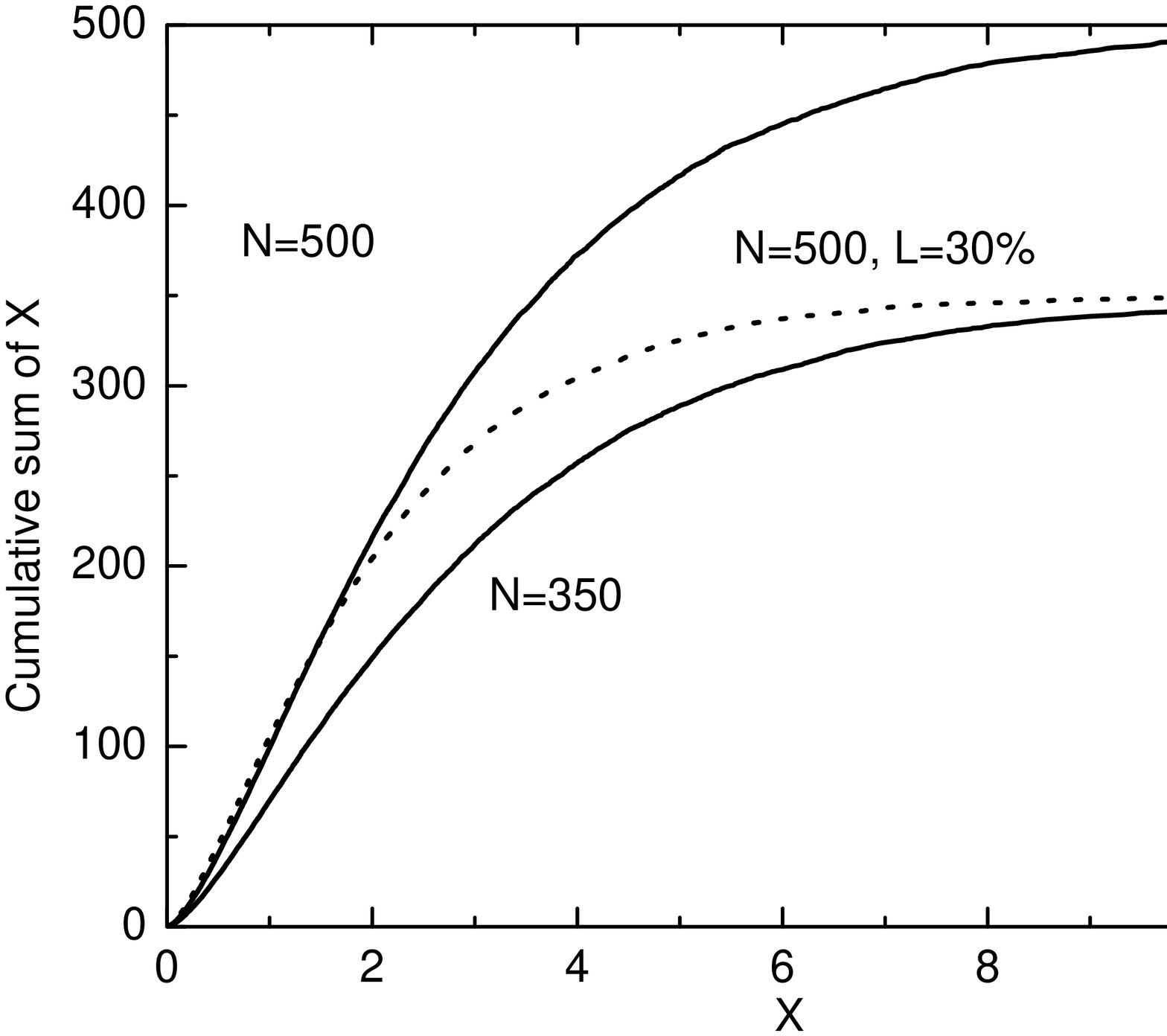} 
\end{center}
\vspace{-5cm}

Fig. 5. Comparison of averaged cumulative sums for equal values of $N$
with thresholds $L=0$ and $L=30\%$. 

\end{figure}
Naturally, function $\delta \psi$ can take into account and other factors
distorting experimental width distribution. This accounting can be performed in
frameworks of both some model approaches and concrete experimental data. 

Modeling of the process of determination of the most probable $D$ value was
performed by approximation of cumulative sums of sets of the random $X$
values for $\chi^2$ distributions with some different $N_t$ values.
Approximately 30\% of their lowest values were excluded from every set
(the threshold -- linearly increasing with number of random value). 

Direct use of equation (2) for determination of the most probable $N_t$ with
high reliability, most probably, is not worth-while without solution of,
as a minimum, two problems:

(a) The set of the precise enough (relative or absolute) dispersion of cumulative
sum for every value of $X$ and 

(b) The guarantied determination of location of absolute minimum of $\chi^2$
corresponding to the desired $N_t$ value.

Although these problems are not irresistible of principle but their solution is not
found up to now.

\section{Conclusion}\hspace*{16pt}
The described and tested method of the reduced neutron widths distributions
analysis allows one to get new of principle information on properties
of neutron resonances. In particular Ц to suppose that the values of the
distribution parameters of their neutron amplitudes can correspond to the
set of several distributions with their different mean values and dispersions.



\begin{thebibliography}{99}
 \bibitem{PT} 
        C.F. Porter and R.G.~Thomas, Phys. Rev. 
        {\bf 104}, 483 (1956)
\bibitem{PEPAN-1972} 
	 V.G. Soloviev, Sov. Phys. Part. Nuc. {\bf 3} (1972) 390.
\bibitem{MalSol}
 	 L.A. Malov, V.G. Solov'ev,  Yad. Phys., {\bf 26(4)} (1977) 729.
\bibitem{BNL}  S.F.  Mughabghab, Neutron Cross Sections
	 BNL-325.  V.  1.  Parts B, edited by Mughabhab S.F., Divideenam M., 
	Holden N.E., N.Y. Academic Press, (1984).
\bibitem{IAEA} 
	http://www-nds.iaea.or.at. 
\bibitem{Prep196}
        A.M.~Sukhovoj and V.~A.~Khitrov,
        Preprint No.  E3-2005-196, JINR (Dubna, 2005).
\bibitem{PEPAN-2006} 
  	A. M. Sukhovoj, V. A. Khitrov,
	Physics of Paricl. and Nuclei, {\bf 37(6)} (2006) 899.
\bibitem{PHAN7210}
 	A. M. Sukhovoj, V. A. Khitrov, W. I. Furman, Physics of Atomic Nuclei,
	{\bf 72} (2009) 1759.
 \bibitem{Appr-k}
 A.M. Sukhovoj, W.I. Furman, V.A. Khitrov,    
Physics of Atomic Nuclei,  {\bf 71(6)}  (2008) 982.
\end{thebibliography}
\end{document}